\documentclass[twocolumn,showpacs,preprintnumbers,amsmath,amssymb,
 aps,
 prb,
floatfix,
 lengthcheck,%
]{revtex4-1}
\usepackage{graphicx}
\usepackage[english]{babel} 
\usepackage[utf8]{inputenc}
\usepackage{hyperref}

\usepackage{xcolor}

\begin{document}

 
\title{Colored delta-\texorpdfstring{$T$}{T} noise in Fractional Quantum Hall liquids} 

\author{K. Iyer, J. Rech, T. Jonckheere,  L. Raymond, B. Gr\'emaud, and T. Martin}
\affiliation{Aix Marseille Univ, Université de Toulon, CNRS, CPT, Marseille, France}

\date{\today}

\begin{abstract}
Photons are emitted or absorbed by a nano-circuit under both equilibrium and non-equilibrium situations. Here, we focus on the non-equilibrium situation arising due to a temperature difference between the leads of a quantum point contact, and study the finite frequency (colored) noise.
We explore this delta-$T$ noise in the finite frequency regime for two systems: conventional conductors described by Fermi liquid scattering theory and the fractional quantum Hall system at Laughlin filling fractions, described by the chiral Luttinger liquid formalism. We study the emission noise, its expansion in the temperature difference (focusing on the quadratic component) as well as the excess emission noise defined with respect to a properly chosen equilibrium situation. The behavior of these quantities are markedly different for the fractional quantum Hall system compared to Fermi liquids, signalling the role of strong correlations. We briefly treat the strong backscattering regime of the fractional quantum Hall liquid, where a behavior closer to the Fermi liquid case is observed. 
\end{abstract} 


\maketitle

\section{Introduction}\label{sec:intro} 

In recent years, the study of non-equilibrium noise in mesoscopic devices has generated new investigations, both on the experimental and on the theoretical side. Indeed, instead of using the standard method to impose a non-equilibrium situation by connecting the device to leads with different voltages and generating so called quantum shot noise, experimentalists have opened the  field of ``delta-$T$ noise’’ by choosing instead to apply a thermal gradient and zero voltage drop to the device. In this situation, provided that electron hole symmetry is respected, a finite zero frequency non equilibrium noise can be measured while the current flowing through the device remains zero.

Voltage bias induced quantum noise\cite{blanter00,martin05,schoelkopf97,dejong96} has always been considered as a crucial diagnosis of quantum transport, providing complementary information about the charge of the current carriers or their statistics. Early theoretical works on delta-$T$ noise suggest that it is also relevant to characterize nanoscopic devices.\cite{lumbroso18,mu19a} In particular, in correlated systems such as quantum Hall devices, delta-$T$ noise in one dimensional correlated systems clearly depends on the dimension of the operators which describe the elementary excitations of the system, suggesting that it could provide information about anyonic statistics.

On the experimental side delta-$T$ noise has been studied in atomic break junctions representing quantum point contacts,\cite{lumbroso18} tunnel junctions,\cite{larocque20} integer quantum Hall effect edge channels,\cite{sivre19} under a weak or a strong temperature bias. Also, it has recently been employed to study the heat transport along the edges.~\cite{melcer22}
On the theoretical side, delta-$T$ charge noise (and in some instance heat noise\cite{eriksson21, tesser23}) has been already studied in a vast variety of  systems, ranging from quantum point contacts/tunnel junctions,\cite{mu19a,mu19b,popoff22} resonant levels or quantum dots in the Kondo regime,\cite{hasegawa21} Fractional Quantum Hall systems,\cite{rech20, Oreg22, rebora22} bosonic systems and quantum spin Hall systems,\cite{gornyi22} and normal metal/superconductor junctions.\cite{zhitlukhina20} All of these studies have focused uniquely on zero frequency noise, the experimental regime where the ``white noise’’ has a weak dependence on the frequency because  this frequency scale is sufficiently high so that 1/f noise can be neglected, but also sufficiently low to avoid specific features associated with the non equilibrium conditions which are imposed on the device.

Voltage induced non-equilibrium noise at high frequency, sometimes dubbed ``colored noise’’,\cite{chamon96, bena07} has been discussed and introduced theoretically about a quarter of a century ago\cite{lesovik97,gavish00}. It was pointed out that its measurement requires a quantum treatment of both the noise detector and the nanoscopic device under study. 
It is therefore considered a subtle quantity because of the necessity to distinguish emission noise, where the nanoscopic device emits microwave photons to the quantum detector, from absorption noise where the detector (which in practice has photon occupations specified by the Bose-Einstein distribution for instance) emits photons which are absorbed by the nanoscopic device. Voltage induced finite frequency (colored) noise
in normal metal junctions is characterized by cusps in the emission and absorption noise located at frequencies corresponding to the Josephson frequency associated with the electron charge. Experimentally, the  measurement of colored noise has for a long time shied away experimentalists because of the inherent difficulties of the measurement scheme, but some successes have been won in superconducting hybrid junctions\cite{deblock03,billangeon06} and with the refinement of experimental detection techniques, recently the Josephson frequency\cite{kapfer19} of fractional quasiparticles of the (Laughlin) fractional quantum Hall effect was measured\cite{bisognin19}, constituting the first finite frequency measurement of noise in a correlated electron system, and an alternative diagnosis of the fractional charge (as compared to the measurement of the Fano factor).   

The question which we want to address in the present work is simple, but  the answer may not be so obvious: what is the frequency spectrum of photons emitted/absorbed from a nanoscopic device when the non-equilibrium condition is imposed solely by a temperature gradient? Does finite frequency delta-$T$ noise have specific signatures which can be tied to the scaling dimension of the operators describing the elementary excitations – and thus their statistics ? Similar questions have been addressed in recent works \cite{hubler23, tesser23} for normal metal leads connected by a quantum dot. As a starting point, we explore the physics of finite frequency delta-$T$ noise in a (normal metal) Fermi liquid system. This will subsequently be used as a benchmark to study finite frequency delta-$T$ noise in the fractional quantum Hall effect, the focus of this article.  

The  paper is organized as follows: in Sec.~\ref{sec2} we introduce the emission and absorption noise, as well as the excess emission noise  and the thermal-like contribution of finite frequency noise; in Sec.~\ref{sec3} we discuss finite  frequency noise for Fermi liquids; in Sec.~\ref{sec4} we focus on the Fractional Quantum Hall effect regime and we conclude in Sec.~\ref{sec_conclusion}. 

\section{Emission, absorption and excess noise}
\label{sec2}

As explained in  the literature,\cite{martin05} when considering finite frequency noise, the quantum nature of the noise detector needs to be described on the same footing as the device under study. There exist typically two coupling schemes between  the two circuits: an inductive coupling scheme\cite{lesovik97,gavish00}, where microwave photons are exchanged between the device and a resonant (LC) circuit, or a capacitive coupling scheme,\cite{aguado00} where photons emitted/absorbed by the device trigger inelastic transitions in a nearby measuring circuit where current is measured. As a result, in full generality, two distinct correlators need to be defined in order to define the physically measured noise. The emission noise describes the spectrum of microwave photons emitted to the (quantum) noise detection device:  
\begin{equation}
S_+ (\omega) = \int_{-\infty}^{+\infty}\mathrm{d}\tau ~\langle \delta I(0) \delta I(\tau)\rangle ~\mathrm{e}^{i\omega \tau}~, 
\end{equation}
where $\delta I(\tau)=I(\tau)-\langle I(\tau)\rangle$ describes the deviation of the current operator from the stationary current $\langle I(\tau)\rangle=\langle I\rangle$.
The absorption noise describes the absorption of microwave photons emitted from the detector.
\begin{equation}
S_- (\omega) = \int_{-\infty}^{+\infty}\mathrm{d}\tau ~\langle \delta I(\tau) \delta I(0)\rangle ~\mathrm{e}^{i\omega \tau}~. 
\end{equation}
They are related by the equation $S^+ (\omega)=S^- (-\omega)$, which allows us to consider only the emission noise from this point onward. We focus on a situation where the two lead reservoir have the same chemical potential, while the left (right) reservoir is at temperature $T_L$ ($T_R$). In this situation, and in the presence of electron/hole symmetry, no net current flows ($\langle I\rangle =0$), but the (non-equilibrium) emission noise $S_+(\omega, T_R, T_L)\neq 0$ depends on both temperatures. 

\begin{figure}[t]
\centering
\includegraphics[width=0.48\textwidth]{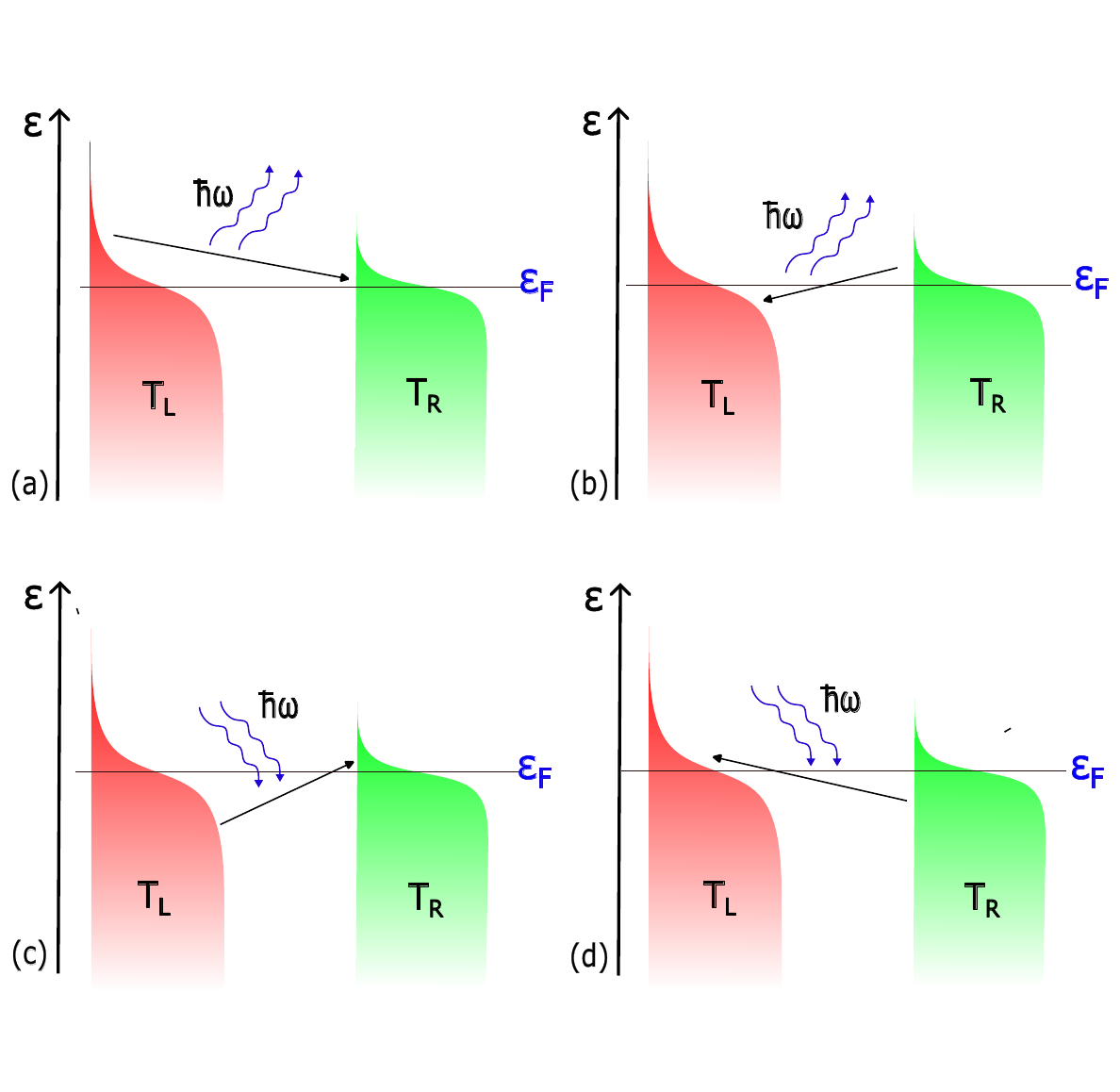}
\caption{The figure depicts the processes giving rise to the emission/absorption noise, that is,  the transmission of electrons from the right Fermi lead to the left lead (or vice versa) accompanied by the emission (subfigures a,b) or absorption (subfigures c,d) of photons of energy $\hbar\omega$. $T_{R/L}$ denote the temperatures of the Fermi leads, where we assume $T_L>T_R$, and $\epsilon_F$ denotes the chemical potential which is the same for both leads. }
    \label{noise_processes}
\end{figure}

We now introduce the ``thermal-like'' contribution to the noise:
\begin{equation}
S_+^{\text{th}}(\omega,T_R,T_L) = \frac{1}{2}S_+(\omega,T_R,T_R)
+\frac{1}{2}S_+(\omega,T_L,T_L)
\label{thermal_noise_defn}
\end{equation}
 which reduces exactly to the finite frequency Johnson-Nyquist thermal equilibrium emission noise when $T_R=T_L$. Following Ref.~\onlinecite{hasegawa21}, it is then convenient to define the excess emission noise according to:
\begin{equation}
\Delta S_+(\omega,T_R,T_L) = S_+(\omega,T_R,T_L)
-S_+^{\text{th}}(\omega,T_R,T_L)
\label{excess_noise_defn}
\end{equation}
where we have subtracted the thermal contributions of both the leads from the emission noise. This quantity is measurable experimentally, and reduces to the sole out-of-equilibrium contribution in the non-interacting regime, even when the transmission probability is energy-dependent.

\section{Fermi liquids}
\label{sec3}

We start by analyzing the general non-equilibrium noise in a system of non-interacting fermions.
Consider a two-terminal phase coherent system composed of fermionic reservoirs separated by a  scattering region specified by a scattering matrix ${\mathcal S}$ (which contains the amplitudes for a particle from reservoir $L$ ($R$) to be transmitted/reflected in reservoir $R$ or $L$; for simplicity, we choose both leads to bear only a single channel).
Each reservoir is described by a Fermi distribution function:
\begin{equation}
f_p (\omega) = \frac{1}{e^{\frac{\hbar \omega}{k_B T_p}} + 1}~.
\end{equation}
where $p$ is  the lead index. 

Noise in such fermionic systems is caused by the transmission of electrons from the left/right lead to the right/left lead, accompanied by the absorption or emission of photons, as depicted in Fig.~\ref{noise_processes}. When considering emission noise, an electron (top left panel) from the  tail of the left (high temperature) Fermi function can lose energy and end up in the vicinity of the Fermi level because there are free states available. This can also happen in reverse (top right panel), but to a lower extent, due to the thermal broadening of the Fermi functions. We emphasize that the latter channel for emission noise is specific to temperature-biased junctions. It is absent for zero-temperature, voltage-biased junctions since there are no states available below the Fermi level. The two lower panels refer to absorption noise processes and both electron transfer processes due to photon absorption are also present for pure voltage-biased junctions. 

Our  starting point is the general formula for finite frequency emission noise:\cite{martin05}
\begin{align}
S_{+}(\omega) =&
\frac{4 e^2}{h} \int dE\sum_{p p'}
\left[\delta_{Lp}\delta_{Lp'}-s^*_{L p}(E) s_{L p'}(E-\hbar\omega)\right]
\nonumber\\
&\hspace{1.8cm}
\times \left[\delta_{Lp'}\delta_{Lp}-s^*_{L p'}(E-\hbar\omega) s_{L p}(E)\right]
\nonumber\\
&\hspace{1.8cm}
\times f_{p}(E) \left[ 1 - f_{p'}(E-\hbar\omega) \right]
~.
\label{eq_corr_bruit_frequence_finie}
\end{align}
where $p$ and $p'$ are lead indices.

The scattering matrix is described by the minimal parametrization:
\begin{equation}
    {\mathcal S}=\begin{pmatrix}
    s_{LL}&s_{RL}\\
    s_{LR}&s_{RR}\\
    \end{pmatrix}
    = \begin{pmatrix}
    i\sqrt{1-{\mathcal T}}& \sqrt{{\mathcal T}}\\
    \sqrt{{\mathcal T}} & i\sqrt{1-{\mathcal T}}
    \end{pmatrix}~,
\end{equation}
where ${\mathcal T}(E)$ is the energy-dependent transmission probability. 

In the context of scattering theory, assuming that the measurement frequency $\omega$ can be neglected in the scattering matrix elements [$s_{pp'}(E-\hbar\omega)\approx s_{pp'}(E)$] the emission noise can be split into thermal (equilibrium) and non-equilibrium (excess) contributions. The thermal-like contribution of the emission noise, given by Eq.~\eqref{thermal_noise_defn} reads, in the context of scattering theory:
\begin{widetext}
\begin{equation}
S_+^{th}(\omega,T_R,T_L)=\frac{2e^2}{h}\int dE~
{\mathcal T}(E)
\left\{f_L(E) \left[ 1-f_L(E-\hbar\omega) \right]+ f_R(E) \left[ 1-f_R(E-\hbar\omega) \right]\right\}
\label{noise_thermal_like}
\end{equation}
\end{widetext}

while the excess emission noise given by Eq.~\eqref{excess_noise_defn} reads
\begin{widetext}
\begin{equation}
    \Delta S_+(\omega,T_R,T_L) = \frac{2e^2}{h}\int dE~\mathcal{T}(E) \left[1-\mathcal{T}(E) \right]
    \left[ f_R(E)-f_L(E) \right] \left[ f_R(E-\hbar\omega)-f_L(E-\hbar\omega) \right]
    \label{STexcessnoise}
\end{equation}
\end{widetext}
which naturally implies that it describes a purely off-equilibrium quantity as $\Delta S_+(\omega,T,T)=0$. Assuming particle-hole symmetry, and using the basic properties of the Fermi distribution, one can prove the following identities:
\begin{eqnarray}
\Delta S_+(\omega,T_R,T_L)&=&\Delta S_+(-\omega,T_R,T_L)\label{parity_rule}\\
\int_{-\infty}^{+\infty}  d\omega~ \Delta S_+(\omega,T_R,T_L)&=&0\label{sum_rule}
\end{eqnarray}
The result of Eq.~\eqref{parity_rule} suggests that the excess emission noise is symmetric with respect to frequency (parity rule) so that it does not distinguish between emission and absorption processes. This symmetry property then allows us to obtain the result of Eq.~\eqref{sum_rule} that the excess emission noise also satisfies a sum rule, where the noise integrated over all frequencies is zero. These features of the excess emission noise will be later examined for fractional quantum Hall liquids. 

In the remainder of this section, we shall assume the transmission coefficient to be constant $\mathcal{T}(E)=\mathcal{T}$, as the scattering theory result will be compared to the (Fermi) filling fraction $\nu=1$ of the quantum Hall effect, where in the wide band limit, Ohm's law is satisfied  and a constant transmission  coefficient is implicit. Note that in this situation, at equilibrium ($T_L=T_R=T$), the thermal contribution of the emission noise has the analytical expression: 
\begin{equation}
S_+^{th}(\omega,T,T)=\frac{4e^2}{h}{\mathcal T}\frac{\hbar \omega }{\exp(\hbar\omega/k_B T)-1}
\label{thermal_noise}
\end{equation}
which, by definition of Eq.~\eqref{thermal_noise_defn}, yields the usual zero frequency Johnson-Nyquist thermal noise as $\omega\to 0$.

\subsection{Small temperature gradient}

We define the temperature difference $\Delta T = T_R - T_L$ and the average temperature $T_\text{avg} = (T_R+T_L)/2$.  Working up to lowest order in the transmission amplitude, we ignore the $\mathcal{T}^2$ term in the non-equilibrium part of the noise for later comparison with the weak backscattering regime of the fractional quantum Hall effect. 

\begin{figure}
    \includegraphics[width=0.48\textwidth]{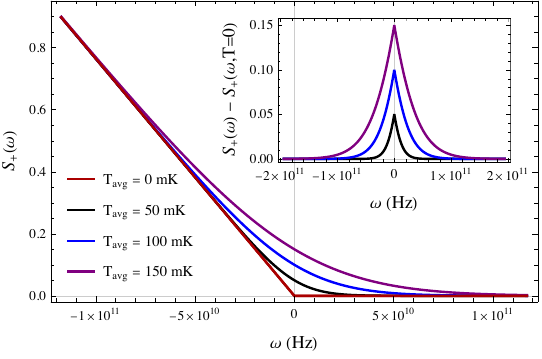}
    \caption{ Emission noise at different $T_{\text{avg}}$, for the regime $\Delta T \ll T_\text{avg}$, for fixed $\Delta T = 10 mK$ ($\Delta T = 0$ for $T_\text{avg} = 0 mK$.) The noise is computed for a transmission $\mathcal{T}=0.01$, and expressed in units of $S_+^{(0)} = e^2 \mathcal{T}/(2 \hbar)$. The decay rate of the spectrum as $\omega$ increases is related to the average temperature of the leads. Higher $T_\text{avg}$ leads to a slower decay which is a consequence of the Fermi distributions broadening. In the inset, we subtracted the zero temperature emission noise from the finite temperature one. }
    \label{emission_noise_fermi}
\end{figure}

The full emission noise $(S_+)$ is plotted in Fig.~\ref{emission_noise_fermi}, for a fixed gradient ($\Delta T = 10 mK$) and several average temperatures.  We note that in the small $\Delta T$ regime, $S_+(\omega, T_R,T_L)$ is almost equal to  $S_+^{\text{th}}(\omega, T_\text{avg}, T_\text{avg})$, given by Eq.~\eqref{thermal_noise}, the difference between the two being only of order $O \left( \frac{\Delta T}{T_\text{avg}}\right)$. Going from the left to the right of the plot, $S_+$ corresponding to different $T_\text{avg}$ are equal for large, negative $\omega$ and decrease linearly. As we get closer to $\omega =0$ $S_+$ still keep decreasing, but the curves corresponding to different $T_\text{avg}$ branch off and the curves with higher $T_\text{avg}$ decay at a slower rate. The temperature-dependent decay continues for $\omega >0$ and eventually, for large, positive $\omega$, all the curves vanish. These features can be understood as a consequence of the thermal broadening of the Fermi distributions, by looking at Fig.~\ref{noise_processes}, where $\omega > 0$ corresponds to subfigures (i), (ii) (emission processes) and $\omega < 0$ to subfigures (iii), (iv) (absorption processes.) A greater number of higher energy states are occupied as $T_\text{avg}$ is increased, hence, for $\omega > 0$, $S_+$ decays slower as a function of frequency until it ultimately vanishes -- corresponding to energies where the state occupation is negligible. Likewise, for $\omega < 0 $, the slower decay of $S_+$ for higher $T_\text{avg}$, can be understood in a similar fashion. For large negative omega, the distinction between Fermi distributions corresponding to different $T_\text{avg}$ is negligible, and the noise is essentially the same, caused by absorption of high-frequency photons by the low energy states. This picture is better understood from the inset of Fig.~\ref{emission_noise_fermi} which shows the difference between the emission noise at a given temperature and the same quantity evaluated at zero temperature. This reflects precisely the change in the occupation of the levels due to a non-zero $T_\text{avg}$.

As pointed out earlier, the non-equilibrium  consequences of the temperature difference are completely masked by the equilibrium thermal noise for $\Delta T \ll T_\text{avg}$. This can also be checked by plotting the emission noise for a fixed $T_\text{avg}$ and different $\Delta T$ where one finds that the curves almost all collapse with the pure equilibrium thermal noise.

\begin{figure}[t]
    \centering
    \includegraphics[width=0.48\textwidth]{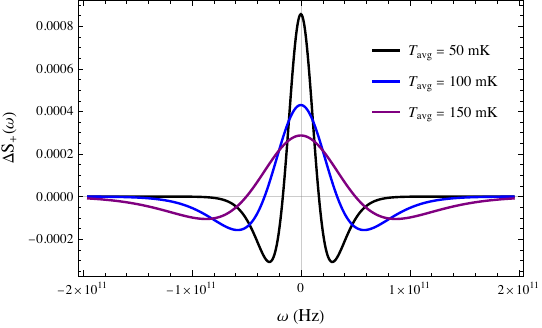}
    \includegraphics[width=0.48\textwidth]{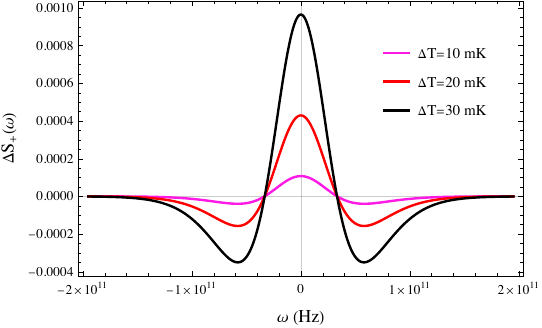}
    \caption{(i) Excess emission noise at a fixed $\Delta T = 20 mK$ and different $T_{\text{avg}}$, (ii) Excess emission noise at the same $T_{\text{avg}} = 100 mK$ but different $\Delta T$ -- both for $\Delta T \ll T_\text{avg}$. The noise is computed for a transmission $\mathcal{T}=0.01$, and expressed in units of $S_+^{(0)} = e^2 \mathcal{T}/(2 \hbar)$. The spread of the excess noise spectrum scales linearly with the average temperature of the leads whereas the magnitude of the excess noise increases quadratically with the temperature difference. }
    \label{excess_noise_fermi}
\end{figure}

This motivates us to look at the \emph{excess emission noise} $(\Delta S_+)$ given by Eq.~\eqref{STexcessnoise}, which has been designed specifically to get rid of the thermal contributions in the non-interacting regime and isolate the non-equilibrium contributions to the noise arising from the temperature difference~\cite{hasegawa21}. $\Delta S_+$ is displayed in Fig.~\ref{excess_noise_fermi}, the top panel showing the excess emission noise for a fixed temperature gradient and several average temperatures, while the bottom panel corresponds to a fixed average temperature but several values of the temperature gradient. In all cases, $\Delta S_+$ is characterized by a central peak at $\omega=0$, where the noise is positive. 
Indeed, for small frequencies, the $\Delta T$-biased system is noisier than the corresponding equilibrium system averaged over the two temperatures. $\Delta S_+$ then decreases gradually, bearing negative values for intermediate positive/negative frequencies, reaching a minimum whose position scales with the average temperature. Negative noise in the intermediate frequency regime suggests that there is \emph{less} noise in the $\Delta T$ non-equilibrium scenario compared to an equilibrium situation of equal temperatures on both the leads. Finally, for large positive or negative frequencies, the excess noise vanishes meaning that the temperature difference does not modify the noise substantially compared to the equilibrium noise in this regime. The change in the sign of $\Delta S_+$ for different frequency regimes can again be understood as a consequence of the difference in the occupation of the left and right Fermi leads. In the delta-$T$ biased regime, for small frequencies, there is a higher number of processes contributing to the noise compared to an equilibrium situation, making the excess noise positive. On the contrary, for intermediate frequencies, there are fewer processes contributing to the noise, giving a negative excess noise.

\begin{figure}[t]
    \includegraphics[width=0.48\textwidth]{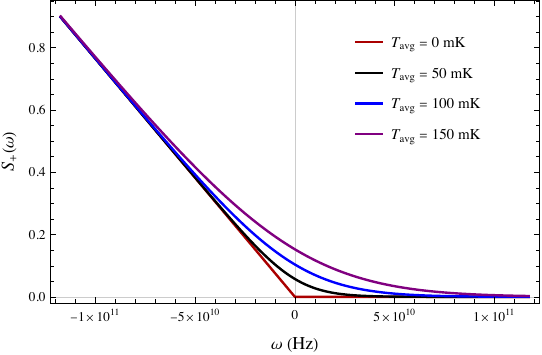}
    \caption{Emission noise at different $T_{\text{avg}}$, for the regime $T_R \ll T_L$ or $\Delta T \sim T_\text{avg}$. The noise is computed for a transmission $\mathcal{T}=0.01$, and expressed in units of $S_+^{(0)} = e^2 \mathcal{T}/(2 \hbar)$. Like in the small $\Delta T$ regime, we find that the noise spectrum decays slower with higher average temperature.}
    \label{emission_noise_fermi_CR}
\end{figure}

\begin{figure}
    \centering
    \includegraphics[width=0.48\textwidth]{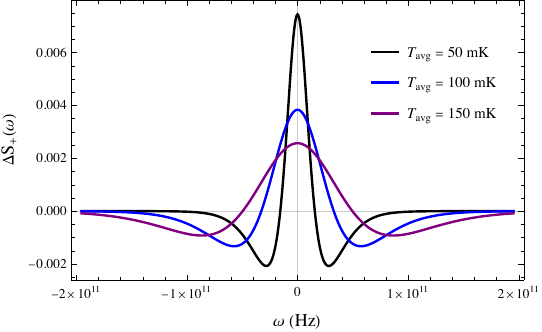}
    \includegraphics[width=0.48\textwidth]{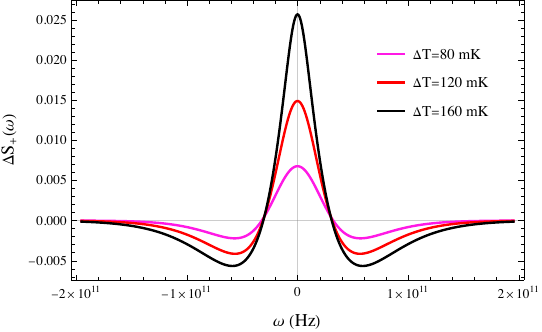}
    \caption{(i) Excess emission noise for a fixed $\Delta T$ (= $60 mK$ and different $T_{\text{avg}}$, (ii) Excess emission noise at the same $T_{\text{avg}} (= 100 mK)$ but different $\Delta T$. The noise is computed for a transmission $\mathcal{T}=0.01$, and expressed in units of $S_+^{(0)} = e^2 \mathcal{T}/(2 \hbar)$. For both the plots we consider $T_L,  T_R$ or $\Delta T \sim T_\text{avg}$. This regime gives a behaviour of the noise spectra quite similar to the small temperature difference regime. }
     \label{excess_noise_fermi_CR}
\end{figure}

As predicted by the parity rule of Eq.~\eqref{parity_rule}, $\Delta S_+$ is an even function of frequency, and the sum  rule Eq.~\eqref{sum_rule} has been checked numerically to be satisfied. We note that for a fixed $\Delta T$, the peak and minima are more pronounced when the average temperature is small. This goes together with an overall spread infrequency which increases linearly as $T_\text{avg}$ increases. However, for a fixed $T_\text{avg}$ and increasing $\Delta T$, while the spread of the spectrum remains the same, the size of the peak in $\Delta S_+$ increases quadratically. It follows that the spread in frequency of the $\Delta S_+$ spectrum seems to be governed by the average temperature, $T_{\text{avg}}$ of the system, while the magnitude of the excess noise, which reflects the \emph{degree of non-equilibriumness} of the noise, seems to be largely dictated by the temperature difference, $\Delta T$. We express this quantitatively as $\Delta S_+(\omega, T_R, T_L) \sim \Delta T^2/T_\text{avg} \mathcal{S} \left( \omega/T_\text{avg}\right)$.

\subsection{Large temperature gradient}

We next consider a non-equilibrium scenario where $T_R \ll T_L$, such that we can essentially consider $T_R \sim 0$. Our consideration of this regime is motivated by its relatively easy experimental accessibility.~\cite{larocque20} 

We again find that the decay rate of the emission noise is controlled strongly by the average temperature of the system (see Fig.~\ref{emission_noise_fermi_CR}). The higher the average temperature, the slower the decay. The excess emission noise in this large temperature difference regime displays a behavior analogous to the small $\Delta T$ case. Again, the spread of the excess noise spectrum depends strongly on the average temperature while the magnitude of the noise is fixed by the temperature difference, as can be seen in Fig.~\ref{excess_noise_fermi_CR}. However, in contrast with the small $\Delta T$ case, $S_+$ may substantially differ from $S_+^\text{th}$ in this regime since for large enough $\Delta T$, the magnitude of $\Delta S_+$ may be comparable to that of $S_+^\text{th}$

\section{Fractional Quantum Hall Effect}
\label{sec4}

\subsection{Luttinger Liquid model}

 We now turn to a Hall bar in the fractional quantum Hall (FQH) regime, with a Laughlin filling factor, ie., $\nu = 1/(2n + 1)~(n \in\mathbb{Z}^+)$. We want to analyse the behaviour of delta-$T$ noise in these systems, which constitutes the central part of this work. FQH systems host edge states that can be described by a chiral Luttinger liquid Hamiltonian given by\cite{martin05, wen95review}
\begin{equation}
    H_0 = \frac{v_F}{4\pi\nu}\int dx\left[ (\partial_x\phi_R)^2 + (\partial_x \phi_L)^2 \right]
\label{CLLHamiltonian}
\end{equation}
where $\phi_{R/L}$ are chiral bosonic fields that describe the right/left moving modes, propagating with velocity $v_F$. The bosonic fields are quantized by the commutation relation $[\phi_{R/L}(x), \phi_{R/L}(y)] = \pm i\pi \text{sgn}(x-y)$ and are related to the quasiparticle operators on the edge through the identity:
\begin{equation}
\psi_{R/L}(x,t) = \frac{U_{R/L}}{\sqrt{2\pi a}}e^{\pm ikx}e^{-i\sqrt{\nu}\phi_{R/L}(x,t)},  
\end{equation} 
where $a$ is a short-distance cutoff, $U_{R/L}$ are the Klein factors, and $k_F$ the Fermi momentum. We further equip the Hall bar with a quantum point contact (QPC), placed at position $x=0$, allowing tunneling between the counter-propagating edges. Working in the weak backscattering regime, where quasiparticles are allowed to tunnel between the edges, we need to add a tunneling term to the Hamiltonian
\begin{equation}
    H_\text{WB}(t) = \Gamma_0 \psi_R^\dagger(0,t) \psi_L(0,t) + \text{H.c.}
\end{equation}
where $\Gamma_0$ is the tunneling amplitude. With this, the tunneling current operator can be calculated to be
\begin{equation}
    I_T(t) = ie^*\Gamma_0 \psi_R^\dagger(0,t) \psi_L(0,t) + \text{H.c.}
\end{equation}
where $e^* = \nu e$ is the quasiparticle charge. 

We compute the delta-$T$ emission noise associated with the backscattering current at the QPC using the Keldysh formalism, to lowest order ($\Gamma_0^2$) in the tunneling amplitude:\cite{martin05} 
\begin{equation}
    S_+(\omega, T_R, T_L) = \left( \frac{e^*\Gamma_0}{\hbar \pi a} \right)^2 \int d\tau~ e^{i\omega \tau} e^{ \nu \mathcal{G}_R(-\tau) + \nu \mathcal{G}_L(-\tau)  }
    \label{luttinger_noise}
\end{equation}
where $T_R, T_L$ are the temperatures at the right- and left-moving edges respectively, $\omega$ is the frequency at which the noise is measured and $\mathcal{G}_{R/L}$ are the finite-temperature bosonic Greens functions of the bosonic fields $\phi_{R/L}$, typical of the chiral Luttinger liquids modelling the FQHE:
\begin{equation}
\mathcal{G}_{R/L}(\tau) = \text{ln} \left[\dfrac{\text{sinh}\left(i\pi \frac{k_B}{\hbar} T_{R/L}\tau_0 \right)}{\text{sinh}\left(\pi \frac{k_B}{\hbar} T_{R/L}(i\tau_0 - \tau) \right)} \right]
\label{chiral_green}
\end{equation}
with $\tau_0 = a/v_F$ being a short time cutoff. For $T_R = T_L = T_\text{avg}$ in Eq.~\eqref{luttinger_noise}, the thermal equilibrium emission noise can be evaluated analytically and is given by
\begin{align}
    S^\text{th}_+(\omega, T_\text{avg}, T_\text{avg}) &= 
    \left( \frac{e^*\Gamma_0}{\hbar \pi a} \right)^2 \tau_0 \left( \frac{2\pi k_B T_\text{avg}}{\hbar}  \tau_0\right)^{2\nu -1} \nonumber \\
    & \times \exp \left(-\frac{\hbar \omega}{2 k_B T_\text{avg}} \right) \frac{\left|\Gamma\left( \nu + \frac{i \hbar \omega}{2 \pi k_B T_\text{avg}}\right)\right|^2}{\Gamma(2\nu)}
\label{luttinger_equilibrium_noise}
\end{align}

\subsection{Small temperature gradient}

\begin{figure}
    \centering
    \includegraphics[width=0.48\textwidth]{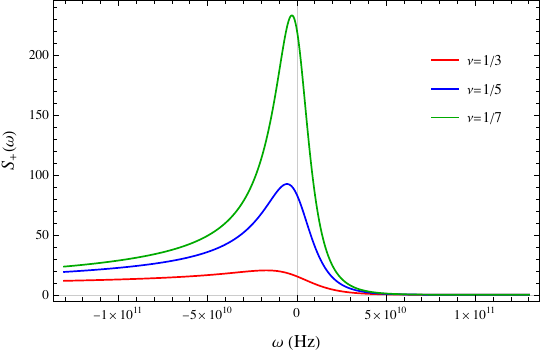}
    \includegraphics[width=0.48\textwidth]{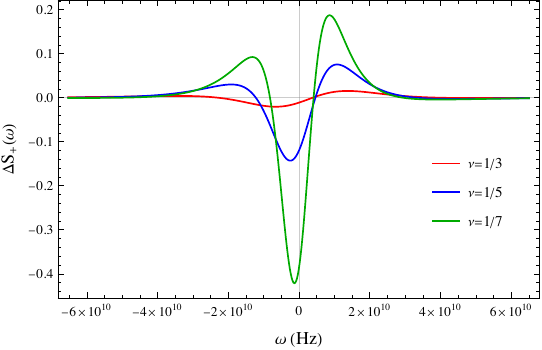}
    \caption{(i) Emission and (ii) excess noise (rescaled) at a QPC comprising two FQH edges in the weak backscattering regime, held at a $10 ~mK$ temperature difference, at an average temperature of $100~ mK.$ The noise is here expressed in units of $\tilde{S}_+^{(0)} = \left( \frac{e^* \Gamma_0}{\pi \hbar v_F} \right)^2 \left(2 \pi \frac{k_B}{\hbar} \right)^{2\nu-1} \frac{\tau_0^{2\nu}}{\Gamma(2\nu)}$ and computed for a small time cutoff $\tau_0$ such that $k_B \tau_0/\hbar = 10^{-5} K^{-1}$. Note that the emission noise looks the same as the equilibrium noise since the consequences of the temperature difference are $10^3$ times smaller than the equilibrium contributions.}
    \label{FQHEemissionexcess}
\end{figure}

We now discuss the properties of delta-$T$ noise in the strongly correlated regime of the Laughlin fractional quantum Hall effect.
The delta-$T$ emission and excess noise, for the weak backscattering regime where anyons tunnel across the QPC, are plotted in Fig.~\ref{FQHEemissionexcess}, for several values of the fractional filling factors $\nu=1/3,1/5,1/7$. 
For the sake of comparison with the Fermi liquid results of the previous section, we use here the same convention of excess emission noise, which was defined in Eq.~\eqref{excess_noise_defn}. Note that, although this definition ensures that, in the non-interacting regime, all thermal contributions are filtered out, leaving only the purely non-equilibrium contributions to the noise, such a cancellation is not guaranteed in the FQH regime and may only be partial.

In the FQH regime, similarly to the Fermi liquid case, the full emission noise ($S_+$) is almost equal to the equilibrium thermal noise ($S_+^\text{th}$), given in Eq.~\eqref{luttinger_equilibrium_noise}. However, the general behavior is quite different from that of the Fermi liquid case, as the emission noise now shows a central asymmetric peak at small negative frequencies, then decreases for large positive/negative frequencies. The sharp decrease for positive frequencies is reminiscent of the Pauli blocking which restricts the emission of photons due to the presence of a Fermi sea. On the other hand, the slow decrease of $S_+$ for negative frequencies has no Fermi liquid equivalent. This behavior at high frequency can be readily understood by considering the asymptotics of Eq.~\eqref{luttinger_equilibrium_noise}. For large, positive frequencies, one has $S^\text{th}_+(\omega\to \infty, T_\text{avg}, T_\text{avg}) \sim \omega^{2\nu -1}\text{exp}\left(-\frac{\hbar \omega}{2 k_B T_\text{avg}}\right)$, thus explaining the rapid exponential decay with frequency. Whereas in the limit of large negative frequency, it reduces to a simple power law in $\omega$ given by $S^\text{th}_+(\omega\to -\infty, T_\text{avg}, T_\text{avg}) \sim \omega^{2\nu -1}$. This power law behavior is directly related to the scaling dimension of the tunneling operator. It has been checked numerically.

Interestingly, the noise spectrum always satisfies the inequality $S_+(-\omega, T_R, T_L) \geq S_+(\omega, T_R, T_L)$, independently of the temperatures of the incoming edge states. This property can be proven exactly in the case of Fermi liquid leads and holds irrespective of the details of the junction or the temperature difference. It amounts to stating that the rate at which the system absorbs energy from the electromagnetic field is always greater than or equal to the rate at which it transfers energy to the field.~\cite{hubler23} This is typically interpreted in terms of processes involving electrons and holes being scattered in the conductor before recombining to emit or absorb the energy of a photon.~\cite{zamoum16} It is quite striking to observe that this generalizes to the case of FQH devices, suggesting a similar interpretation based on quasiparticle-quasihole pairs.

Contrary to the Fermi liquid case, the excess emission noise $\Delta S_+$ is asymmetric in frequency for nontrivial Laughlin fractions, which constitutes another example of the role of electronic correlations in the FQH regime. This breaks the parity rule of Eq.~\eqref{parity_rule}, departing from the Fermi liquid picture. However, and quite importantly, the excess emission noise still satisfies the sum rule of Eq.~\eqref{sum_rule}, despite its asymmetry in frequency. This can be readily understood upon integrating the expression of Eq.~\eqref{luttinger_noise} over the whole frequency range, noticing from Eq.~\eqref{chiral_green} that $\mathcal{G}_{R/L}(\tau=0)=0$, so that the integrated emission noise reduces to a constant, independently of the temperature of the leads. This result for the sum rule has also been checked numerically. 
 
\begin{figure}[t]
    \centering
    \includegraphics[width=0.48\textwidth]{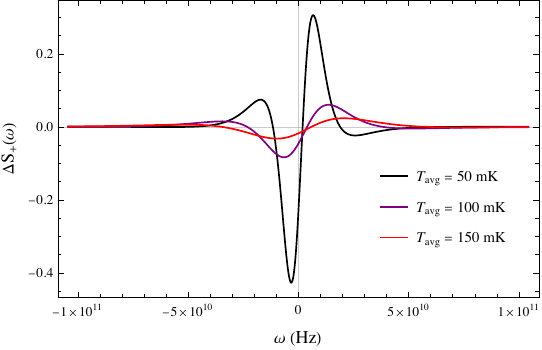}
    \includegraphics[width=0.48\textwidth]{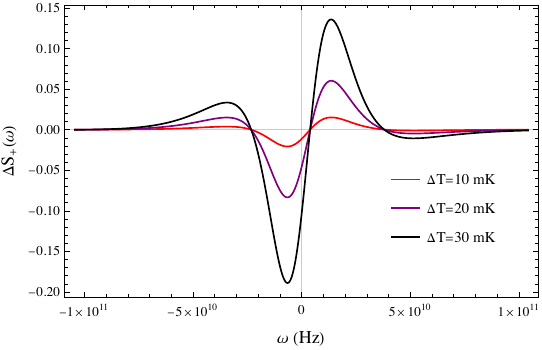}
    \caption{(i) Excess emission noise at a fixed $\Delta T = 20 mK$ and different $T_{\text{avg}}$ for FQH $\nu = 1/3$ edges, (ii) Excess emission noise at the same $T_{\text{avg}} = 100 mK$ but different $\Delta T$ for FQH $\nu = 1/3$ edges -- both for $\Delta T \ll T_\text{avg}$. The noise is here expressed in units of $\tilde{S}_+^{(0)} = \left( \frac{e^* \Gamma_0}{\pi \hbar v_F} \right)^2 \left(2 \pi \frac{k_B}{\hbar} \right)^{2\nu-1} \frac{\tau_0^{2\nu}}{\Gamma(2\nu)}$ and computed for a small time cutoff $\tau_0$ such that $k_B \tau_0/\hbar = 10^{-5} K^{-1}$. Similar to the Fermi liquid regime, the spread of the excess noise spectrum is dictated by the average temperature of the leads whereas the magnitude of the excess noise is fixed largely by the temperature difference.}
    \label{Luttinger_one_third_excess_temperature_dependence}
\end{figure}

We now look at the behaviour of $\Delta S_+$ focusing on the filling factor $\nu = 1/3$ FQH, first as a function of the average temperature ($T_\text{avg}$) for a fixed temperature difference ($\Delta T$), and then as a function of $\Delta T$ for a fixed $T_\text{avg}$. The results are displayed in Fig.~\ref{Luttinger_one_third_excess_temperature_dependence}. First, we find that the ($\Delta T, T_\text{avg}$) dependence of $\Delta S_+$ for $\nu = 1/3$ FQH, is largely similar to that of the Fermi liquid regime. Indeed, even in this strongly correlated system, we find that the spread in frequency of the noise spectrum is a function of the average temperature of the entire system, whereas the magnitude of the excess noise depends primarily on the temperature difference between the two FQH edges. Other filling fractions display the same behavior (not shown.) This behavior can be described quantitatively by an expression of the form $\Delta S_+\left( \omega, T_R, T_L) \right) \sim T^{2\nu -3}\Delta T^2 \mathcal{S} \left( \omega/T_{\text{avg}} \right)$, for small $\Delta T$. As $\Delta T$ increases, the higher-order contributions become more important and we observe deviation from this behavior.

\begin{figure}[t]
    \centering
    \includegraphics[width=0.48\textwidth]{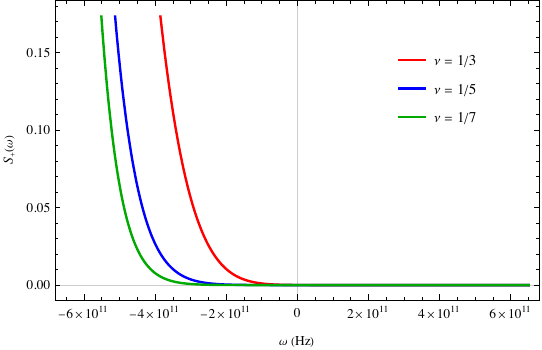}
    \includegraphics[width=0.48\textwidth]{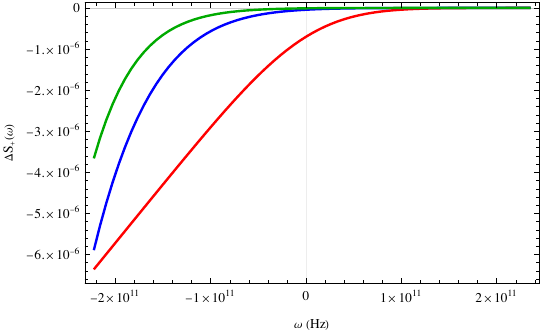}
    \caption{(i) Emission and (ii) excess noise (rescaled) at a QPC comprising two FQH edges in the strong backscattering regime, held at a $10 ~mK$ temperature difference, at an average temperature of $100~ mK.$ The noise is here expressed in units of $\tilde{S}_+^{(0)} = \left( \frac{e^* \Gamma_0}{\pi \hbar v_F} \right)^2 \left(2 \pi \frac{k_B}{\hbar} \right)^{2\nu-1} \frac{\tau_0^{2\nu}}{\Gamma(2\nu)}$ and computed for a small time cutoff $\tau_0$ such that $k_B \tau_0/\hbar = 10^{-5} K^{-1}$.}
    \label{SBS}
\end{figure}

Comparing the results of Fig.~\ref{excess_noise_fermi_CR} to those of Fig.~\ref{Luttinger_one_third_excess_temperature_dependence}, one first notices an overall sign flip of the excess emission noise with a similar-looking structure involving three extrema. The central zero-frequency peak of the Fermi liquid case is now shifted toward negative frequency signaling a strong reduction of the absorption. While the side peaks are also present, they differ from the Fermi liquid case in that they are no longer symmetric, occurring at frequencies that are seemingly unrelated, with a bigger amplitude at positive frequencies, corresponding to a stronger enhancement of the emission. This behavior is qualitatively reminiscent of the one observed for a resonant level asymmetrically coupled to Fermi liquids~\cite{hubler23}, and as such could be related to the nontrivial energy dependence of the scattering at the QPC.

For completeness, we now consider the strong backscattering regime of the FQHE, where electrons, instead of anyons tunnel across the QPC. This regime can be accessed by invoking the standard duality properties of the chiral Luttinger liquid description of the edge states, which amounts to simply replacing $\nu \to 1/\nu$ and $e^* \to e$ in Eq.~\eqref{luttinger_noise}. We show the emission and excess noise for several values of the filling factors in Fig.~\ref{SBS}. 

The emission noise $S_+$ in the strong backscattering regime is closer to the one of Fermi liquids than that of the FQH weak backscattering regime, rapidly decaying for $\omega > 0$ and growing for $\omega < 0$, without any significant features. For negative frequencies, the emission noise now grows as a power law of $|\omega|^{{2}/{\nu}-1}$, as opposed to the simple linear-in-frequency behavior observed in the Fermi liquid case. Again, a simple interpretation of this behavior of the emission noise is hard to come by, but one may point out that this is associated with the scaling dimension of the tunneling operator, which now involves electrons rather than anyons.

The excess noise $\Delta S_+$ in the strong backscattering regime is rather featureless, only showing a reduction of the absorption. Note that a careful examination of the excess noise at extremely high frequency does show a peculiar behavior. This is actually an artifact of the calculation as it happens for frequencies beyond the scale set by the cutoff of the theory, namely $\omega \gg v_F/a$. These results are unphysical and only signal a breakdown of the chiral Luttinger liquid description at such high energies.


Lastly, we note that the Luttinger liquid results map back exactly to the Fermi liquid results if one sets $\nu = 1$, as expected. This has been dealt with analytically in Appendix \ref{appendixa}.

\subsection{Temperature gradient expansion}

Unfortunately, Eq.~\eqref{luttinger_noise} is not analytically tractable in its full form, motivating us to treat it perturbatively in the small temperature gradient limit. Following the zero frequency delta-$T$ noise analysis of Ref.~\onlinecite{rech20}, starting from $T_{R/L} = T_{\text{avg}}\pm \frac{\Delta T}{2}$ where $\Delta T \ll T_{\text{avg}}$, we expand the exponentiated Green's function perturbatively up to second order in $\Delta T/2$, giving us (we assume $\hbar = k_B =1$ in this section to declutter the equations)
\begin{equation}
     S_+(\omega, T_R, T_L) = S_0(\omega,T_{\text{avg}})\left[ 1 + \left(\frac{\Delta T}{2T_{\text{avg}}} \right)^2 C_2(\omega,T_{\text{avg}}) \right]
\end{equation}
where
\begin{equation}
    S_0(\omega, T_{\text{avg}}) = \left( \frac{e^*\Gamma_0}{\pi a} \right)^2 \int d\tau e^{i\omega \tau}\left[\dfrac{\text{sinh}\left(i\pi T_{\text{avg}}\tau_0 \right)}{\text{sinh}\left(\pi T_{\text{avg}}(i\tau_0 + \tau) \right)} \right]^{2\nu}
\label{S0}
\end{equation}
\begin{widetext}
and
\begin{equation}
    C_2(\omega, T_{\text{avg}}) = \frac{1}{S_0(T_{\text{avg}},\omega)}\left( \frac{e^*\Gamma_0}{\pi a} \right)^2 \int d\tau~ e^{i\omega \tau}\left[\dfrac{\text{sinh}\left(i\pi T_{\text{avg}}\tau_0 \right)}{\text{sinh}\left(\pi T_{\text{avg}}(i\tau_0 + \tau) \right)} \right]^{2\nu} \left[ \frac{\nu(\pi(i\tau_0 +\tau))^2 }{\text{sinh}(\pi T_{\text{avg}}(i\tau_0 +\tau))} - \frac{\nu(i\pi \tau_0)^2 }{\text{sinh}(i\pi \tau_0 T_{\text{avg}})}\right]
\label{C2}
\end{equation}
\end{widetext}
Here, $S_0(\omega, T_\text{avg}) \equiv S^\text{th}_+(\omega, T_\text{avg}, T_\text{avg})$ is just the equilibrium thermal noise, already evaluated in Eq.~\eqref{luttinger_equilibrium_noise}. Both the integrals of Eq.~\eqref{C2} can also be evaluated analytically. The details of the calculation  are summarized in Appendix \ref{appendixb}.  The result for $C_2$ then reads:
\begin{widetext}


\begin{align}
   C_2\left(\frac{\omega}{2\pi T_\text{avg}}\right) &=  \nu \left[ -1 + \frac{\big|\nu + \frac{i\omega}{2\pi T_\text{avg}} \big|^2}{2\nu (2\nu +1)} \left( \pi^2 +4 \pi \text{Im} \left[\psi\left( \nu + 1 + \frac{i\omega}{2\pi T_{\text{avg}}} \right) \right] \right. \right. \nonumber \\
& \qquad +4 \left. \left. \left\{ \text{Im} \left[\psi\left( \nu + 1 + \frac{i\omega}{2\pi T_{\text{avg}}} \right) \right]  \right\}^2 - 2 \text{Re} \left[\psi'\left( \nu + 1 + \frac{i\omega}{2\pi T_{\text{avg}}} \right) \right]  \right) \right]
\label{C2_eval}
\end{align}
\end{widetext}
where $\psi$ is the digamma function and prime indicates a derivative. The $C_2$ coefficient in Eq.~\eqref{C2_eval} is obtained directly from an expansion of the emission noise, following in that respect the convention adopted in earlier works.~\cite{rech20, gornyi22} This corresponds to a slightly different definition of the excess noise compared to the one used so far and defined in Eq.~\eqref{excess_noise_defn}. It corresponds to an excess noise where the reference noise is chosen to be the equilibrium noise at the average temperature, i.e. $C_2 = \left[S_+(\omega, T_R, T_L) - S_+(\omega, T_\text{avg})\right]/S_+(\omega, T_\text{avg})$. While one could equally introduce an equivalent coefficient by expanding in $\Delta T$ the excess noise $\Delta S_+$ defined in Eq.~\eqref{excess_noise_defn}, this is merely a matter of convention, and ultimately allows to highlight different properties. Here, we resort to the present choice since it readily distinguishes the weak and strong backscattering regime,~\cite{rech20} which is not so clear with other conventions.

\begin{figure}[tb]
    \centering
    \includegraphics[width=0.48\textwidth]{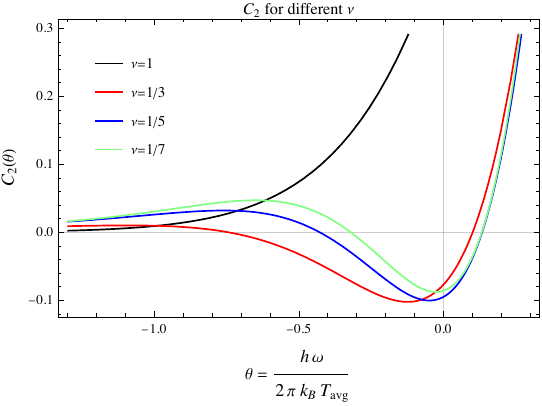}
\caption{The $C_2$ coefficient for FQH edges, plotted with respect to the dimensionless quantity $\theta = \frac{\hbar\omega}{2\pi k_B T_\text{avg}}$, with the QPC operating in the weak-backscattering regime. The coefficient displays dips to negative values for Laughlin fractions. }
    \label{C2coefficients}
\end{figure}

Interestingly, it turns out that $C_2(\omega, T_{\text{avg}})$ actually does not depend separately on frequency and temperature, but rather in a combined way, being a function of the ratio $\omega/T_\text{avg}$. The behavior of this $C_2$ coefficient, which encodes the relevant ''non-equilibrium" information, is plotted in Fig.~\ref{C2coefficients} as a function of  $\theta = \hbar \omega/(2 \pi k_B T_\text{avg})$ in the case of weak backscattering at the QPC.
There is a clear distinction between the behavior for the Laughlin fractions and the one for the trivial integer case. While for $\nu=1$, the $C_2$ coefficient increases monotonically, it displays a dip, crossing into negative values for frequencies close to zero for the Laughlin fractions. The value of the minimum is only marginally affected by the filling factor (within the Laughlin sequence), however the range of frequency over which $C_2 < 0$ is $\nu$-dependent and shrinks with the filling factor. In all cases, the $C_2$ coefficient grows as a power-law at high frequency, nevertheless the contribution to the emission noise is washed out by the exponential decay of the equilibrium thermal noise.

\begin{figure}[tb]
    \centering
    \includegraphics[width=0.48\textwidth]{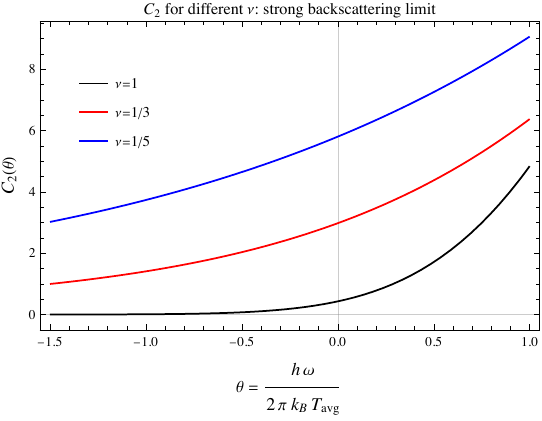}
    \caption{The $C_2$ coefficient for FQH edges, plotted with respect to the dimensionless quantity $\theta = \frac{\hbar\omega}{2\pi k_B T_\text{avg}}$, with the QPC operating in the strong-backscattering regime. Here, the coefficient remains positive throughout.}
    \label{C2coefficientsstrongBS}
\end{figure}

In the strong backscattering regime, which is accessed by making use of the duality properties and simply replacing $\nu \to 1/\nu$ in Eq.~\eqref{C2_eval}, we find that the curves for Laughlin FQH show a strong resemblance to the $\nu = 1$ curve, monotonically increasing as a function of $\omega/T_\text{avg}$ with no dips to negative values as shown in Fig.~\ref{C2coefficientsstrongBS}. 

\section{Conclusions}
\label{sec_conclusion}

This work dealt with the finite frequency spectrum of photons emitted from a thermal gradient generated non equilibrium transport in both Fermi and quantum Hall junctions. The finite frequency noise was characterized here by the emission noise as well as with the excess emission noise, which has solely non equilibrium origins in the Fermi picture as thermal noises of each reservoirs are subtracted. For electron-hole symmetric Fermi junctions, the Landauer-B\"uttiker formalism can be employed, and 
the excess noise does not distinguish between emission and absorption processes as it is an even function of frequency. The excess emission noise of Fermi liquid thus has a central positive peak, and changes sign at moderate frequencies, acquires a minimum and then vanishes to zero. The height of the peak is controlled by the temperature difference and its width is determined by the average temperature.  

For a QPC in the fractional quantum Hall regime, we employed the chiral Luttinger liquid theory to compute  in the weak backscattering regime the emission and excess noise when both edges have  different temperatures. We started with the weak backscattering regime which is dominated by quasiparticle tunneling where new physics is expected. While the emission noise vanishes for positive frequencies, it also does for negative frequencies which departs strongly from the Fermi liquid picture. The emission noise has a central, asymmetric peak for small negative frequencies. The excess noise contains a minimum for small negative frequencies, in sharp contrast with the Fermi liquid case, and it is also asymmetric. The excess noise can be explored by varying  both the average temperature and the temperature gradient. 

The emission noise in the strong back-scattering regime, where only electrons can tunnel between the two semi infinite Hall fluids, resembles strongly the Fermi liquid case (it decays as positive frequencies and grows for negative frequencies) but it follows a Luttinger liquid power law (rather than the linear behavior predicted by Fermi liquid theory) at negative frequencies.

It seemed judicious to follow Ref.~\onlinecite{rech20} and explicitly perform a thermal gradient expansion of the emission noise (in the weak backscattering regime) to characterize the coefficient $C_2$ of the quadratic term in the gradient, which was obtained analytically as a function of the ratio between the frequency and the average temperature. $C_2$ is negative and has a minimum for small negative frequencies (in accordance with the zero frequency result). It grows for positive frequencies and decays to zero for negative frequencies. $C_2$ plotted as a function of frequency allows to further point out the differences with Fermi liquid theory. In the strong backscattering regime $C_2$ behaves roughly as in the Fermi liquid case, monotonically increasing, with no minima of negative contributions. 

This work does open the path to the investigation of finite frequency noise in mesoscopic systems driven out of equilibrium by a thermal gradient. While the regime of Fermi liquids was used here primarily as a benchmark and a point of comparison, we believe that our study of the strongly correlated regime of the fractional Hall effect deserves attention, as on many instances, departures from the Fermi liquid picture are observed.    

\acknowledgments

This work received support from the French government under the France 2030 investment plan, as part of the Initiative d'Excellence d'Aix-Marseille Université - A*MIDEX.
We acknowledge support from the institutes IPhU (AMX-19-IET-008) and AMUtech (AMX-19-IET-01X).

\bibliography{mainFQH_biblio}

\onecolumngrid
\appendix

\section{Connection between Scattering Theory and the Integer Quantum Hall Effect}
 \label{appendixa}

In this appendix, we show that the emission noise of a QPC between two IQH edges expressed in the langauge of Luttinger liquids is equivalent to the noise in the transmission of electrons between Fermi liquids described by scattering theory. We start from the emission noise of Luttinger liquids with $\nu = 1$
\begin{equation}
    S_+(\omega, T_R, T_L) = \left( \frac{e^*\Gamma_0}{\pi a} \right)^2 \int d\tau~ e^{i\omega \tau} \text{exp}\left[  \mathcal{G}_R(-\tau) +  \mathcal{G}_L(-\tau)  \right]
\end{equation}

where the Greens function $\mathcal{G}_{R/L}(\tau)$ is given by
\begin{equation}
    \mathcal{G}_{R/L}(\tau) = -\text{ln}\left[ \frac{\text{sinh}\left[i\pi T_{R/L}\tau_0 \right]}{\text{sinh}\left[\pi T_{R/L}\left(i\tau_0 + \tau \right)\right]} \right]
\end{equation}
To proceed, we define
\begin{equation}
    D_{R/L}(\epsilon) = \int d\tau e^{i\epsilon\tau}D_{R/L}(\tau )
    \label{fouriertransformdefn}
\end{equation}
where 
\begin{equation}
    D_{R/L}(\tau) = \frac{1}{2\pi a}~\frac{\text{sinh}\left[i\pi T_{R/L}\tau_0 \right]}{\text{sinh}\left[\pi T_{R/L}\left(i\tau_0 + \tau \right)\right]}
\end{equation}
Now
\begin{equation}
    S_+(\omega, T_R,T_L) = (2e\Gamma_0)^2 \int d\tau ~e^{i\omega \tau}~D_{R}(\tau)~D_L(\tau)
\end{equation}
Inverting the Fourier transform relation in Eq.~\eqref{fouriertransformdefn}
\begin{align}
    S_+(\omega, T_R,T_L) &= (2e\Gamma_0)^2 \int d\tau ~e^{i\omega \tau} \int \frac{d\omega_R}{2\pi}e^{-i \omega_R\tau}D_R(\omega_R)  \int \frac{d\omega_L}{2\pi}e^{-i \omega_L\tau}D_L(\omega_L) \nonumber \\
    &= (2e\Gamma_0)^2 \int d\tau \frac{d\omega_R}{2\pi}\frac{d\omega_L}{2\pi}~2\pi\delta(\omega-\omega_R-\omega_L)~D_R(\omega_R)~D_L(\omega_L)
    \label{noiseexpression}
\end{align}
We now want to calculate $D_{R/L}(\omega_{R/L})$
\begin{equation}
    D_{R/L}(\epsilon) = \frac{1}{2\pi a}\int d\tau~ e^{i\epsilon\tau}\left[\frac{\text{sinh}\left(i\pi T_{R/L}\tau_0 \right)}{\text{sinh}\left[\pi T_{R/L}\left(i\tau_0 + \tau \right) \right]} \right]
\end{equation}
Setting $\pi T_{R/L}\tau_0 \equiv \alpha$ and $\pi T_{R/L}\tau \equiv u$, we can coax this integrand into the following expression
\begin{align}
   D_{R/L}(\epsilon) &= \frac{1}{\pi T_{R/L}} \frac{1}{2\pi a} e^{\epsilon\alpha/\pi T_{R/L}}\int du~e^{-\frac{i\epsilon(u-i\alpha)}{T_{R/L}}}~\frac{\text{sinh}(i\alpha)}{\text{sinh}(i\alpha - u)} \nonumber \\
   &= \frac{1}{\pi T_{R/L}} \frac{1}{2\pi a} e^{\epsilon\alpha/\pi T_{R/L}}~A_{1/2}\left(\frac{\epsilon}{T_{R/L}} \right)
\end{align}
where
\begin{equation}
    A_{\nu}(z) = \frac{1}{2}(\text{sin}\alpha)^{2\nu}e^{-\frac{\pi}{2}z}\frac{ \left| \Gamma\left(\nu + iz/2 \right)\right|^2}{\Gamma(2\nu)}
\end{equation}
Giving us, 
\begin{equation}
    D_{R/L}(\epsilon) = \frac{1}{\pi T_{R/L}} \frac{1}{2\pi a} e^{\epsilon\alpha/\pi T_{R/L}}~\text{sin}\alpha~e^{-\frac{\pi}{2}z} \left| \Gamma\left(\frac{1}{2} + i\frac{z}{2} \right)\right|^2
\end{equation}
Using the identities $\left| \Gamma\left(\frac{1}{2} + i\frac{z}{2} \right)\right| = \sqrt{\frac{\pi}{\text{cosh}(\pi z)}}$, and then $\tau_0 = a/v_F$ and taking the limit $a\longrightarrow 0, $ we end up with
\begin{align}
    D_{R/L}(\epsilon) &= \frac{\pi}{2v_F}\frac{1}{1 + e^{\pi \epsilon/T_{R/L}}} \nonumber \\
    &\equiv \frac{\pi}{2v_F}f_{R/L}(\pi\epsilon)
\end{align}
Plugging this back into Eq.~\eqref{noiseexpression}, we have
\begin{equation}
    S_+(\omega, T_R,T_L) = \frac{1}{2}\left(\frac{e\Gamma_0}{v_F}\right)^2 \int dE~f_R(E)~f_L(\omega-E)
\end{equation}
Playing around with this result, we end up with the following expression for the noise
\begin{align}
    S_+(\omega, T_R,T_L) &= e^2 \left(\frac{\Gamma_0}{2v_F} \right)^2 \int dE~\left\{ f_L(E) \left[1-f_L(E-\omega) \right] \right. \nonumber \\ 
    & \qquad \left. + f_R(E) \left[1-f_R(E-\omega)\right]+ \left[f_R(E-\omega) - f_L(E-\omega)\right] \left[f_R(E) - f_L(E) \right] \right\}
\end{align}
which is equal to the scattering theory noise expression if we make the identification $\pi \left(\frac{\Gamma_0}{2v_F} \right)^2 \equiv \mathcal{T}$.

\section{\texorpdfstring{$C_2$}{C2} analytics}
\label{appendixb}

Here, we briefly go over the evaluation of the integrals in Eqs.~\eqref{S0} and \eqref{C2}. The perturbatively expanded noise is given by

\begin{align}
     S_+(\omega, T_R, T_L) &= \left( \frac{e^*\Gamma_0}{\pi a} \right)^2 \int d\tau~ e^{i\omega \tau}\left[\dfrac{\text{sinh}\left(i\pi T_{\text{avg}}\tau_0 \right)}{\text{sinh}\left(\pi T_{\text{avg}}(i\tau_0 + \tau) \right)} \right]^{2\nu} \nonumber \\
& \qquad \times \left\{ 1 + \left(\frac{\Delta T}{2T_{\text{avg}}} \right)^2  \left[ \frac{\nu(\pi(i\tau_0 +\tau))^2 }{\text{sinh}(\pi T_{\text{avg}}(i\tau_0 +\tau))} - \frac{\nu(i\pi \tau_0)^2 }{\text{sinh}(i\pi \tau_0 T_{\text{avg}})}\right] \right\}
 \label{appendixnoise}
\end{align}

Absorbing the constants into the variables, the key integral to be evaluated takes the following generic form

\begin{equation}
A_{\nu}(z) = \int du~e^{-iz(u-i\alpha)}\left[\frac{\text{sinh}(i\alpha)}{\text{sinh}(i\alpha - u)} \right]^{2\nu}
\label{integraltobeevaluted}
\end{equation}

Rewriting the hyperbolic sine in terms of exponentials, the integral can be recast as
\begin{equation}
    A_{\nu}(z) = (1- e^{-2i\alpha})^{2\nu}e^{-z\alpha}\int du \frac{e^{-(2\nu+iz)u}}{\left[ e^{-2u} + e^{i(\pi-2\alpha)}\right]^{2\nu}}
    \label{GRform}
\end{equation}
This integral can be evaluated using Ref.~\onlinecite{gradshteyn14} (3.314) which gives us
\begin{equation}
    \int dx \frac{e^{-\tilde{\mu} x}}{\left[ e^{-x/\tilde{\gamma}} + e^{\tilde{\beta}/\tilde{\gamma}}\right]^{\tilde{\nu}}} = \tilde{\gamma}\text{exp}\left[ \tilde{\beta}\left(\tilde{\mu} - \frac{\tilde{\nu}}{\tilde{\gamma}} \right)\right]\frac{\Gamma\left(\tilde{\nu}-\tilde{\gamma}\tilde{\mu} \right)\Gamma\left( \tilde{\gamma}\tilde{\mu}\right)}{\Gamma\left( \tilde{\nu}\right)}
\end{equation}
provided the conditions $\text{Re}\left( \frac{\tilde{\nu}}{\tilde{\gamma}}\right) > \text{Re}~\tilde{\mu} > 0$ and $|\text{Im}~\tilde{\beta}| < \pi~\text{Re}~\tilde{\gamma}$ are satisfied. Here, $\Gamma(x)$ is the Euler-Gamma function. For the integral in Eq.~\eqref{GRform}, we can identify $\tilde{\beta} = i(\pi/2 - \alpha)$, $\tilde{\gamma} = 1/2$, $\tilde{\nu} = 2\nu$ and $\tilde{\mu} = 2\nu + i z$, which satisfies all the conditions. We then finally have

\begin{equation}
    A_\nu(z) = \frac{1}{2}(2\sin\alpha)^{2\nu}e^{-\frac{\pi}{2}z}\frac{\left| \Gamma\left( \nu + i\frac{z}{2}\right)\right|^2}{\Gamma(2\nu)}
    \label{GRresult}
\end{equation}
The first integral in Eq.~\eqref{appendixnoise} can be evaluated, after trivial manipulations, directly using Eq.~\eqref{GRresult}. The second term in the second integral can be evaluated similarly. The first term in the second integral of Eq.~\eqref{GRresult} is a bit more involved and is related to the second derivative of Eq.~\eqref{GRresult} with respect to $z$. This can be seen from Eq.~\eqref{integraltobeevaluted}, where a $z$-derivative will bring down a factor $(u-i\alpha)$. The second derivative of Eq.~\eqref{GRresult} can be expressed as
\begin{align}
        \partial_z^2A_\nu(z) &= \frac{1}{2}\frac{(2\sin\alpha)^{2\nu}}{\Gamma(2\nu)}e^{-\frac{\pi}{2}z}\left| \Gamma\left( \nu + i\frac{z}{2}\right) \right|^2 \frac{1}{4}\Bigg\{\pi^2 - \left[ \psi\left(\nu + i\frac{z}{2} \right) - \psi\left(\nu - i\frac{z}{2} \right) \right]^2 \nonumber \\ 
        & -2i\pi\left[ \psi\left( \nu + i\frac{z}{2}\right)-\psi\left( \nu - i\frac{z}{2}\right)\right] -\left[ \psi'\left( \nu + i\frac{z}{2}\right)+\psi'\left( \nu - i\frac{z}{2}\right)\right]  \Bigg\}
    \label{GRresultderivative}
\end{align}
where $\psi(z)$ is the digamma function and $\psi'(z)$ its $z$-derivative. Finally, using Eqs.~\eqref{GRresult} and \eqref{GRresultderivative}, we can express the full noise, in the limit $\tau_0 \longrightarrow 0$ as
\begin{align}
S_+(\omega, T_R, T_L) &= \left( \frac{e^*\Gamma_0}{\pi a} \right)^2\tau_0 \left[2\pi\tau_0 T_\text{avg}\right]^{2\nu -1}e^{-\frac{\omega}{2T_\text{avg}}} \frac{\left| \Gamma\left( \nu + \frac{i\omega}{2\pi T_\text{avg}}\right)\right|^2}{\Gamma(2\nu)} \nonumber \\
 & \times \left\{ 1+ \left( \frac{\Delta T}{2 T_\text{avg}} \right)^2  \nu \left[ -1 + \frac{\big|\nu + \frac{i\omega}{2\pi T_\text{avg}} \big|^2}{2\nu (2\nu +1)} \left( \pi^2 + 4\pi \text{Im} \left[\psi\left( \nu + 1 + \frac{i\omega}{2\pi T_{\text{avg}}} \right) \right] \right. \right. \right. \nonumber \\
& \qquad \left. \left. \left. + 4 \left\{ \text{Im} \left[ \psi\left( \nu + 1 + \frac{i\omega}{2\pi T_{\text{avg}}} \right) \right]\right\}^2 - 2\text{Re} \left[\psi'\left( \nu + 1 + \frac{i\omega}{2\pi T_{\text{avg}}} \right)  \right]  \right) \right] \right\}
\end{align}

From this, we can extract the coefficient $C_2$ which isolates the non-equilibrium contributions in second order of the temperature difference
\begin{align}
   C_2\left(\frac{\omega}{2\pi T_\text{avg}}\right) & = \nu \left[ -1 + \frac{\big|\nu + \frac{i\omega}{2\pi T_\text{avg}} \big|^2}{2\nu (2\nu +1)} \left( \pi^2 + 4\pi \text{Im} \left[\psi\left( \nu + 1 + \frac{i\omega}{2\pi T_{\text{avg}}} \right) \right] \right. \right.  \nonumber \\
& \qquad  \left. \left. + 4 \left\{ \text{Im} \left[ \psi\left( \nu + 1 + \frac{i\omega}{2\pi T_{\text{avg}}} \right) \right]\right\}^2 - 2\text{Re} \left[\psi'\left( \nu + 1 + \frac{i\omega}{2\pi T_{\text{avg}}} \right)  \right]  \right) \right] 
\end{align}

\end{document}